\documentclass[conference]{IEEEtran}
\usepackage{graphicx}
\usepackage[noadjust]{cite}
\usepackage[acronym]{glossaries} 

\begin{document}

\title{Wireless Sensor Network for Internet of Things}

\author{\IEEEauthorblockN{Nacer Khalil}
\IEEEauthorblockA{School of Science and Engineering\\
Al Akhawayn University In Ifrane\\
Email: n.khalil@aui.ma}
\and
\IEEEauthorblockN{Mohamed Riduan Abid}
\IEEEauthorblockA{School of Science and Engineering\\
Al Akhawayn University In Ifrane\\
Email: r.abid@aui.ma}
\and
\IEEEauthorblockN{Driss Benhaddou}
\IEEEauthorblockA{School of Engineering and Technology\\
University of Houston\\
Email: dbenhaddou@uh.edu}
\and
\IEEEauthorblockN{Michael Gerndt}
\IEEEauthorblockA{Institut fuer Informatik\\
Technical University of Munich\\
Email: gerndt@in.tum.de}}
\maketitle

\begin{abstract}
The Internet is smoothly migrating from an Internet of people towards an Internet of Things (IoT). By 2020, it is expected to have 50 billion things connected to the Internet. However, such a migration induces a strong level of complexity when handling interoperability between the heterogeneous Internet things, e.g., RFIDs (Radio Frequency Identification), mobile handheld devices, and wireless sensors. In this context, a couple of standards have been already set, e.g., IPv6, 6LoWPAN (IPv6 over Low power Wireless Personal Area Networks), and M2M (Machine to Machine communications). In this paper, we focus on the integration of wireless sensor networks into IoT, and shed further light on the subtleties of such integration. We present a real-world test bed deployment where wireless sensors are used to control electrical appliances in a smart building. Encountered problems are highlighted and suitable solutions are presented.
\end{abstract}

\section{Introduction}
The Internet of Things (IoT) is smoothly migrating from an Internet of people towards an Internet of Things. According to Cisco \cite{ref1}, 50 billion things will be connected to the Internet in 2020, thus overshadowing the data generated by humans. This is limited by the birth rate: in 2020, it is expected to have 8 billion people worldwide \cite{ref2}. The things to be connected to the Internet largely vary in terms of characteristics. This ranges from very small and static devices (e.g., RFIDs) to large and mobile devices (e.g., vehicles). Such heterogeneity induces complexity and stipulates the presence of an advanced middleware that can mask this heterogeneity and promote transparency. In particular, Wireless Sensor Networks (WSNs) are connecting things to the Internet through a gateway that interfaces the WSN to the Internet. Unlike other networks, WSNs have the particular characteristic of collecting sensed data (temperature, motion, pressure, fire detection, Voltage/current, etc) and forwarding it to the gateway through a one-way communication protocol. Even though most WSN protocols were not designed for two-way communications, they should also be able to receive information and send it to the sensors (as a form of a command for instance), and react on behalf of the commander/user, e.g., automating home appliances. \\
IoT will integrate rich set of applications into the Internet, e.g., automation, weather sensing, and Smart Grids (SGs). The latter is one of the most promising IoT applications. In SGs, Wireless Sensors are used to measure and keep track of energy consumption and production in order to optimize energy usage. \\
In general, Internet things communicate by producing and consuming information and execute “smart” algorithms to interact intelligently with other things in the Internet. Besides, Internet Protocol Version 6 (IPv6) is used to uniquely identify the things in the Internet. To enable the integration of WNS in the IoT, there are two key points that should be added to the relevant protocols: First, the IPv6 over Low power Wireless Personal Area Networks (6LoWPAN) protocol should be implemented and deployed in Wireless Sensor Networks (WSNs); Second, Machine to Machine communications (M2M) protocols \cite{ref3} need to be standardized. \\
In this paper, we deploy a wireless sensor network (WSN) test bed and use 6LoWPAN to leverage wireless sensors as Internet end-with a two- way communication capability. The deployed tested is composed of a WSN, a middleware, and a mobile client for smart home energy monitoring and control. Data is collected from the motes within the WSN and communicated to the middle-ware. The mobile client is able to monitor and visualize the sensed data and control appliances remotely. The main two contributions of this paper are:
\begin{enumerate}
\item Identifying the challenges of deploying IPv6 over 6LoWPAN, and ways to interface with IPv4 networks. The paper presents the performance of the deployed network in terms of delay in different segments of the network.
\item Identifying the challenges of deploying a two-way communication between the wireless sensors and the Internet users, and implementing in the WSN.
\end{enumerate}
The rest of the paper is organized as follows: Section II presents related work and background. Section III describes the system architecture. In section IV, the deployment of the system is highlighted and section V presents relevant experiments evaluating the system. The paper is concluded by section VI.
\section{Background and Literature Review}
IoT is a new Internet paradigm based on the fact that there will be much more things than humans connected to the Internet. This means that machines/things will be able to communicate autonomously without the need to interact with human beings, thus rendering them into becoming the major entity generating data in the Internet. Currently, there are already over 12.5 billion things connected to the Internet \cite{ref1} and they will surpass humans in terms of the data they generate. In IoT, M2M will be the main communication standard between the Internet things \cite{ref3}. \\
Besides, ubiquitous and pervasive computing are key technologies significantly contributing to the advent of IoT. They bring computing all the way to physical objects which can communicate in the Internet by producing, consuming, and computing information, through RFID, mobile computing, and WSNs among other technologies \cite{ref4,ref5}. RFID tags bear electronic identification data of different physical objects (e.g., goods, cars, and even wearable sensors), and can even used to identify people. RFIDs consume very little energy by reflecting signals received from RFID readers. On the other hand, mobile and handheld devices (e.g., smartphone and PDAs) are changing the way we access and interact with things in the Internet, and is rendering the Internet into a ubiquitous service. Along with cloud computing, the capabilities of these devices will be further boosted by providing storage and computing power in the cloud. \\
WSNs are a prevalent instance of ubiquitous computing that enables small things to connect to the Internet. The sensory data will make a significant portion of the information flowing in the Internet. In particular, Smart Grids are one of the applications where different parts (things) of the grid (e.g., smart meters) communicate in order to optimize energy consumption as well as energy management in the Grid. SGs are heterogeneous by nature as it feeds power to different consumers (Homes, commercial buildings, factories, etc.) and therefore use heterogeneous technologies such as WiMax, WiFi, Zigbee, WSN, 6LoWPAN, M2M and IP Multimedia Subsystem (IMS) \cite{ref6}. \\
Zigbee is one of the technologies used in WSNs and is being adopted as a standard in SG for home area networks to connect appliances, equipment, and producers of energy such as solar panels to communicate information. IPv6, as being part of the wireless sensor network, bring numerous advantages. However, there are challenges that had to be addressed for IPv6 to be implemented on top of Zigbee, namely fragmentation, frame size, addressing, security \cite{ref7}, and IPv4/IPv6 translation. This paper introduces a real test bed that includes the whole TCP/IP protocol implemented by Berkeley Low-power IP stack (BLIP) \cite{ref7} and that takes into consideration most of those issues. The test bed implements the two-way communication as needed by smart grids and measures the performance of such a system. El Kouche et. al \cite{ref8} investigates the widely used WSNs architectures and technologies and highlights the most suitable architectures for WSN deployment into IoT . In [5], authors present the requirements for deploying an IoT gateway, and propose architecture for the corresponding system to be deployed in the gateway. A similar architecture to what is presented in this paper uses Global System for Mobile Communications (GSM) to communicate information \cite{ref9}. \\
From the architectural point of view, integrating SGs into IoT imposes the stringent need of addressing heterogeneity. An IoT gateway system based on Zigbee and GPRS protocols helps partly in dealing with the heterogeneity problem and therefore enables the WSNs to communicate with the mobile telecommunication network \cite{ref10}. Another solution to the heterogeneity problem is proposed with a new light-weight web service transport protocol called Lean Transport Protocol (LTP) \cite{ref11} that allows transparent exchange of web service messages between all kinds of devices. This protocol is platform-independent and uses low-energy communication. Other researchers claim that the major source of heterogeneity arises from the fact that there are different types of WSN devices (e.g. Micaz, Mica2, and Telosb) that do not use the same standards [9]. A proposition has been made to migrate WSN communication towards an ”all-IP” mode. This would eliminate most of the heterogeneity. A relevant architecture is sketched, and is capable of converting all the WSNs, new and legacy, to support IPv6 \cite{ref12}. \\
In order to make the smallest devices connected to the Internet, 6LoWPAN has been used for this purpose. 6LoWPAN is based on the idea that all things should support the TCP/IP protocol stack and thus join the IoT. In order to build the TCP/IP protocol stack in these devices, multiple aspects of IP need to be addressed, basically IP Maximum Transmission unit (MTU) should be fixed at 1280 Bytes whereas in the Zigbee MTU is only 127 Bytes. This means that IPv6 packets cannot be encapsulated within Zigbee frames. Another issue is related to the addressing with the 128 bits address; in 6LoWPAN, IPv6 addressing is performed hierarchically. The main purpose behind is to identify the packet’s destination network ID before forwarding it to the network. These were just two instances of a large set of issues that 6LoWPAN solves in order to enable the low- power devices to join IoT. TinyOS \cite{ref13}, which is a common operating system for WSNs, comes with a lightweight implementation of 6LoWPAN called BLIP. \\
This project makes use of BLIP to provide the TCP/IP protocol stack to the WSN. 6LoWPAN is used at different parts of the system and more details about these parts will be provided in the system architecture section.
\section{System Architecture}
The proposed system for integrating WSNs into IoT is composed of four essential blocks:
\begin{itemize}
\item Wireless Sensor Network (WSN)
\item Gateway Server
\item Middle-ware
\item Mobile client
\end{itemize}

\begin{figure*}[ht]
\centering
\includegraphics[width=\textwidth,height=7cm]{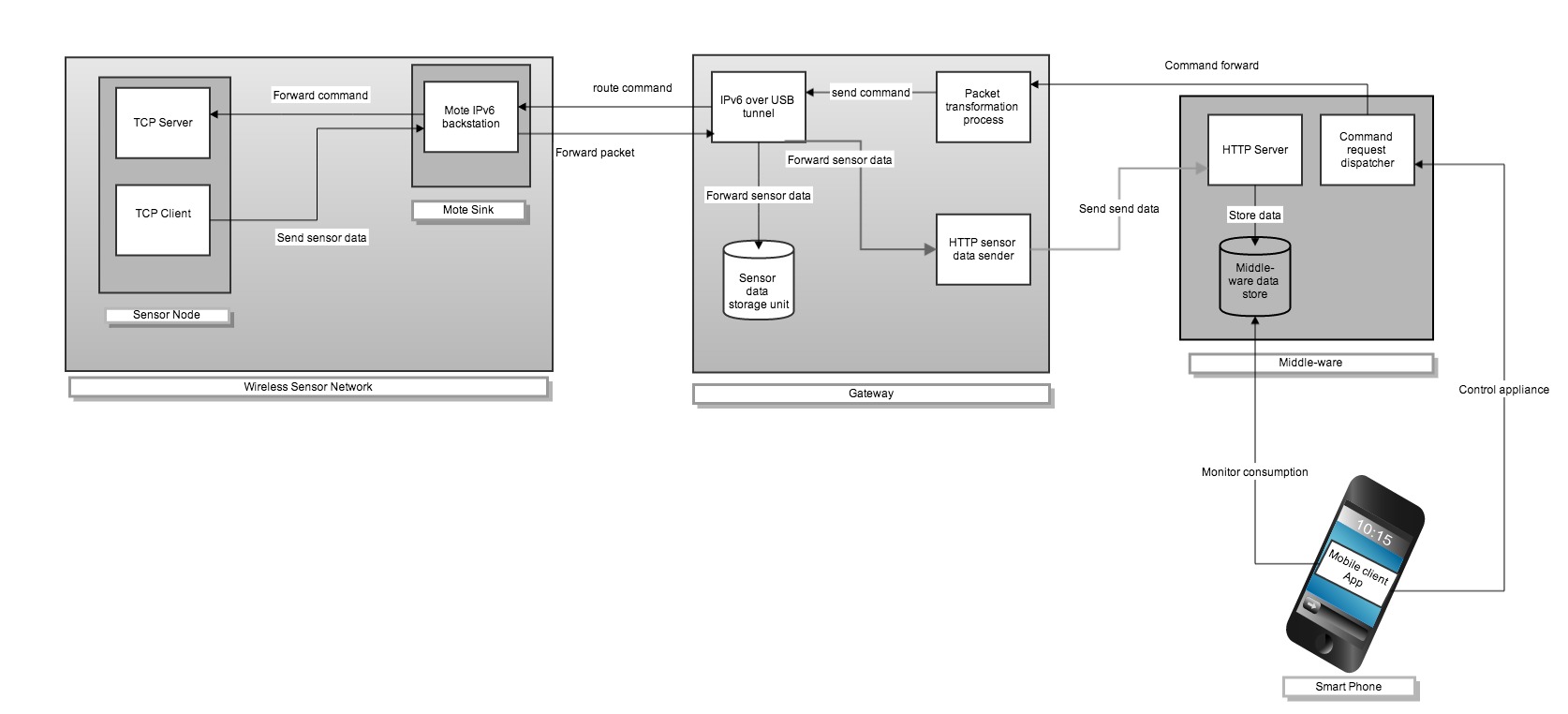}
\caption{General Architecture of the System}
\label{fig:gen_architecture}
\end{figure*}

\begin{figure}[htbp]
\centering
\includegraphics[scale=0.45]{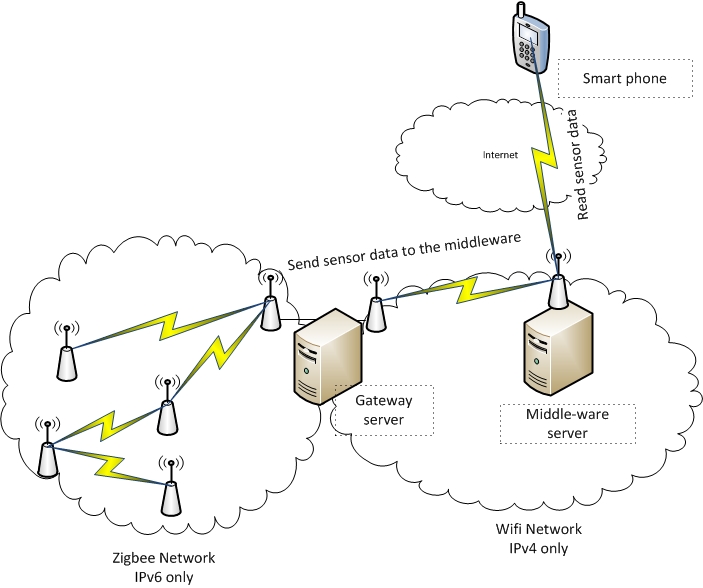}
\caption{Network Diagram}
\label{fig:network_diagram}
\end{figure}

The WSN uses Zigbee as the communication medium and uses IPv6 in the network layer. However, the communication between the gateway server, the middle-ware, and the mobile client is based on IPv4 over Wi-Fi. This architecture enables any device within the system to communicate with any other device independently of the communication medium used (e.g., Zigbee or Wi-Fi) or the network protocol used (e.g., IPv4 or IPv6). In Figure \ref{fig:gen_architecture}, the system architecture is presented. It depicts the four main components of the system along with the relevant subcomponent. This figure also shows the communication flow between the different components of the system. \\
Figure \ref{fig:network_diagram} presents the deployed network diagram, and depicts the different components of the system as well as the interconnections that exist between these different components. \\
 \subsection{Wireless Sensor Network}
The WSNs test-bed is composed of seven motes of type Crossbow MPR2600 \cite{ref14}. From the network topology perspective, the WSN is a multi-hop mesh network that uses the Ad hoc On-Demand Distance Vector (AODV) routing protocol \cite{ref15}. It is an ad-hoc network, whereby motes can be placed anywhere, without a preset topology, as long as there is at least one wireless link for communication. These communication links are created and refreshed dynamically between different motes of the WSN provided that their frames can reach the destination. In addition to the seven motes in the test-bed, there is an additional mote that plays the role of a sink connecting the WSN motes to the gateway server machine. The connection between the sink and the gateway is based on Universal Serial Bus (USB) connection. 
\subsection{Gateway Server}
The gateway server is a key component in the system. It extracts and sniffs Wi-Fi frames, transforms them into Zigbee frames by replacing the appropriate frames’ headers, and forwards them to the sink. In the other direction, the gateway server receives Zigbee frames containing IP packets. These latter get encapsulated in a USB frame and then extracted at the level of the Gateway server to fit in a Wi-Fi frame. \\
The gateway server is also responsible for receiving IPv4 packets and transforming them into IPv6 and vice versa. Besides, it has other functionalities such as receiving sensor data from the WSN and forwarding them to the middle-ware. In case the link between the gateway server and the middle-ware is lost, the gateway server stores the received data in a temporary data store and communicates this data once the link is up again.
\subsection{Middle-ware}
The middle-ware is a software component that is used to mask the heterogeneity in the system, and thus rendering it transparent to external users. The middle-ware also provides automation mechanisms in order to control and reduce the energy consumption. The main features are the ability to receive data, filter it, transform and store it in a coherent fashion in order to use it smartly in order to reduce consumption. In addition, the middle-ware provides an interface to end users via a set of web services that enable them to access all needed information (e.g., real-time and periodic consumption levels), and issue commands to control the appliances through the WSN.
\subsection{Mobile Client}
The mobile client application is an application deployed on Android phones that enables users to access the real-time energy consumption at their homes. Besides, it remotely controls the appliances by turning them On and Off. The mobile client, when wanting to turn On or Off an appliance, sends a command directly to the mote responsible for controlling the appliance and addresses the mote using its ”virtual” IPv4 address. The latter is a virtual one since only IPv6 addresses are supported. A virtual IPv4 address is reserved and assigned for each mote and the translation is made at the gateway level. \\
Now that all components have been introduced, the data flow of the information is to be explained. As it was stated, any component in the system can communicate with any other independently of the data link layer technology or network layer technology.
\subsection{Data Flows}
One of the main goals of this paper is to build a two-way communication between the client and sensor nodes. \\
Figure 3 depicts the data flow diagram corresponding to a mobile user sending a command to the WSN. The mobile client is connected to a Wi-Fi network that uses IPv4 whereas the WSN uses IPv6. Therefore, there should be a process that controls, tracks and transforms the incoming and outgoing packets. The client starts by sending an IPv4 packet to the virtual IPv4 address of the mote. Afterwards, the gateway receives it, translates the virtual IPv4 address into the real IPv6 address of the mote by setting as source address the virtual IPv6 address of the mobile client. The new IPv6 packet is created, carrying the payload coming from the original packet. This new IPv6 packet is forwarded to the wireless sensor network using an IPv6-over- USB tunnel that encapsulates the packet into a USB frame and communicates it to the mote sink. The latter extracts the IPv6 packet from the USB frame and encapsulates it into a Zigbee frame. Once the Zigbee frame arrives to the destination mote, the TCP
datagram is extracted and passed to the TCP server port in the mote that reads the message and executes it by turning On/Off the appliance using I2C (Inter- Integrated Circuit) \cite{ref16}.\\
\begin{figure}[htbp]
\centering
\includegraphics[scale=2.5]{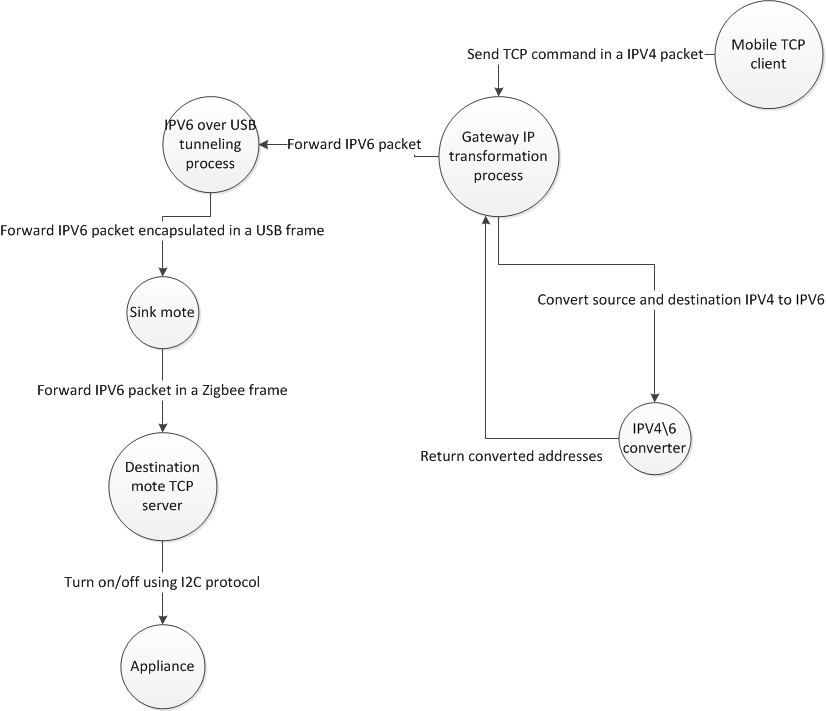}
\caption{Data flow diagram for the mobile client sending On/Off commands}
\label{fig:data_flow_command}
\end{figure}
In the other direction, the mote sends periodically sensory data to the middle-ware. The relevant communication passes through several steps, which are depicted in figure 4:The mote periodically reads sensory data from the sensor, transforms the data and communicates it. To send it, the mote client connects to a TCP server hosted at the gateway server. An IPv6 packet is encapsulated in a Zigbee frame that is forwarded to the mote sink that extracts the IPV6 packet and encapsulates it into a USB frame and then forwards it to the gateway where the TCP datagram is extracted. Once the sensory data is at the gateway, it is communicated to the middle-ware. If the link is down, the sensory data is temporarily stored in a database hosted in the gateway. Once the link is up, all the stored data is sent to the middle- ware and cleared from the database.
\begin{figure}[htbp]
\centering
\includegraphics[scale=2.5]{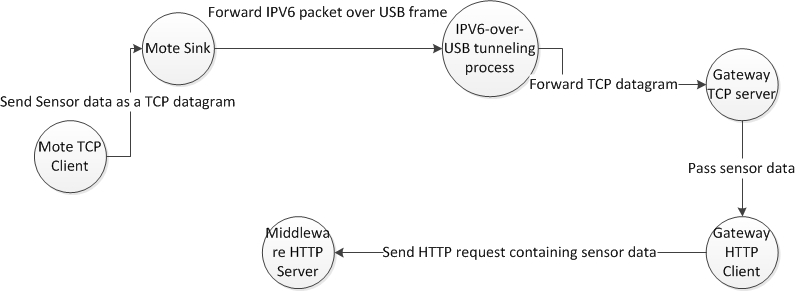}
\caption{Data flow diagram depicting the sending of sensory data}
\label{fig:data_flow_sensing}
\end{figure}
\section{System Deployment}
To meet the constraints of a system capable of providing two-way communication between any host and any mote in the WSN, the following components have been deployed:
\begin{itemize}
\item Mote Programming
\item Mote sink Packet Forwarding
\item IPv4/IPv6 Gateway
\item Network Gateway Sensor Data Server
\end{itemize}
Next sections highlight these components.
\subsection{Mote Programming}
Each mote within the WSN network is equipped with an electric current transformer that is attached to the data acquisition board through which data is read and transformed to the appropriate format and sent to the network host. Besides, the appliance is attached to the mote through the relay pins existing within the data acquisition board in the mote. In other words, the mote can control the electricity going to each mote and can allow or block it. This means that one can control the appliance by using some of the functionalities provided by the mote. From the mote’s perspective there are two parts that are implemented within its TinyOS program. A TCP server that is used to receive On/Off requests in order to control the mote and a TCP client used to send sensor data. Once the program is installed, an IPV6 address is passed to the installation routine in order to assign a static IPv6 address to the mote in which the program is cross-compiled and installed. The TCP server and the TCP client work in parallel as each one’s traffic is handled separately:
\subsubsection{TCP Server}
It is an important component in the mote’s program. To control the appliance, one must connect to the TCP server and send requests. As a mote may control more than one appliance, we identify the appliance by a unique ID and send a zero to turn Off or one to turn On.
\subsubsection{TCP Client}
It serves as means to send sensor data to the gateway in a reliable way. Once the mote is turned On, the client connects to the gateway TCP. The consumption data is then sensed periodically (once per second) and sent to the TCP server who deals processes the sensor data.
\subsection{Mote Sink Packet Forwarding}
The mote sink packet forwarding module is a special program installed within a mote that is equipped with a USB port that plays the role of a network interface card. The mote is attached to the gateway station and has the module within it. In addition, it communicates with the gateway using USB protocol. In the gateway station, the network interface module is an IPv6 over USB tunnel. This means that IPv6 packet destined to the sink are encapsulated within a USB frame, and once it arrives to the sink, the IPv6 packet is extracted and forwarded to the destination mote holding that IPv6 address. The other way around is fairly similar, when a mote wants to send an IPv6 packet to the outside world, the mote creates the packet, sends it to the sink that forwards it by encapsulating it into a USB frame.
\subsection{IPv4/IPv6 Gateway}
This is the most crucial component in the system. It addresses the “gatewaying” issue between IPv4 and IPv6 networks, i.e., the Internet and the WSN. The WSN network supports only IPv6 while other components such as the middle-ware and the mobile client do not necessary have an IPv6 address, but we still want all the components to communicate independently of the IP technology to be used. To do so, we have created a network packet transformation program. This program basically converts IPv4 to IPv6 and vice versa. To do so, it assigns virtual IPv4 addresses to IPv6 address holders and IPv6 address to IPv4 address holders. With such a program, each player in the network has both an IPv4 and an IPv6; still, it is aware of only the one that is assigned to it. The other ”virtual” address is known only at the level of the program installed at the gateway station between the WSN and the outside world. \\
The flow of information works as follows: when a station wants to send requests to a mote, it sends an IPv4 packet holding the request to the mote. This \textit{packet transformation program} that will extract the TCP datagram, create a new IPv6 packet specifying the source address as the virtual address of the host and the destination as the real address of the mote. Afterwards, the TCP datagram is appended to the newly created packet and sent to the mote sink packet forwarding component that is seen by the gateway as a network interface card. Still, this leads to a complicated issue that needs to be handled separately. \\
The issue consists of the fact that the gateway program should keep track of the request responses in order to forward them correctly to the destination. To solve such a problem, an algorithm has been created whose sole role is to mechanically compute the IPv6 address of the host based on its IPv4 and vice versa. This algorithm is based on a mapping function whose primary feature is bijectivity. This means that any IPv4 address is uniquely mapped to one and only one IPv6 adress and vice versa. Thus, whenever a request is coming, the source and destination addresses will be converted using this algorithm, hence avoiding the whole request response tracking part.
\section{Evaluation}
To evaluate the system, we tracked the extent to which the system is able to operate reliably, and offering an acceptable level of performance. Two experiments were conducted to measure the system’s performance. \\
\begin{figure}[htbp]
\centering
\includegraphics[scale=0.62]{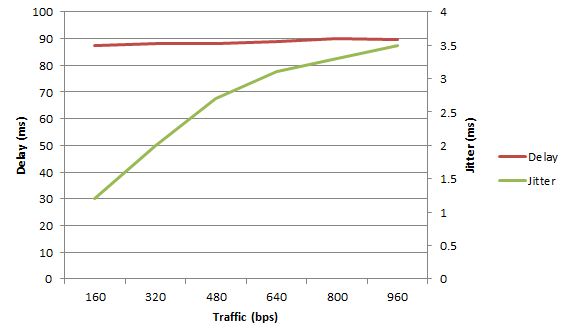}
\caption{Delay and Jitter variation with increasing traffic}
\label{fig:data_flow_command}
\end{figure}

In the first experiment, the behaviour of the system is recorded, where each mote reads sensory data every second and generates traffic in the WSN. We measure the delay and observe its variance regarding the traffic intensity. The routing protocol in principle gives more priority to routing control packets rather than data ones. Therefore, this priority might affect the network’s delay. In Figure 5, we clearly notice that the number of motes in the network does not significantly affect the average delay. On the other hand, the number of motes in the network significantly affects the jitter. The jitter is more sensitive to the change in traffic because there are time intervals where the network’s load is higher than other times which makes the jitter grow and keeping the delay constant. \\
In the second experiment, we measure the contribution of the Gateway Packet Transformation process to the overall communication delay. In other words, how much delay will will be added when adding the packet transformation process?The results present the average delay and jitter computed over the elapsed time starting from the sniffing of the packet in the Gateway Packet Transformation process to the transformation and sending to the recipient. This was carried over 200 packets that were sniffed and transformed by the process. \\
\begin{figure}[htbp]
\centering
\includegraphics[scale=0.72]{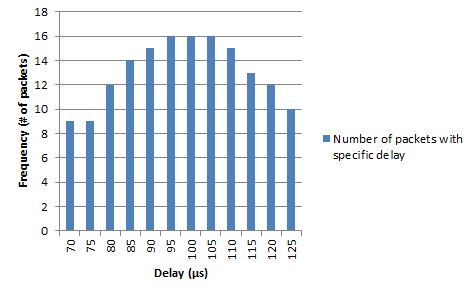}
\caption{Delay Frequency Histogram}
\label{fig:data_flow_command}
\end{figure}
The transformation process’s elapsed time is measured in microseconds. The average delay is on average 100 microseconds whereas the jitter is around 30 microseconds. This means that the process’s time varies between a few microseconds to at most 150 microseconds. In addition, depending on the machine’s load, the distribution of the delay frequency is shown in the histogram depicted in figure 6. From this figure one can conclude that the delay is normally distributed. In addition, this experiment shows that the Gateway Packet Transformation process does not significantly contribute to the overall delay.
\section{Conclusion}
In this paper, we presented the subtleties of integrating wireless sensors networks into the Internet in order to control electrical appliances. We delineated the architecture for deploying a real- world testbed. The presented architecture is simple and can be easily adopted for similar deployments. We highlighted relevant problems mainly IPv4 to IPv6 gatewaying. \\
As a future work, we intend to further research the middleware system component to support heterogeneous wireless sensor motes, and thus not to limit deployment to specific motes, e.g., TinyOS ones.
\mbox{}
\nocite{*}
\bibliographystyle{IEEEtran} 
\bibliography{IEEEabrv,myrefs}

\begin{thebibliography}{10}
\providecommand{\url}[1]{#1}
\csname url@samestyle\endcsname
\providecommand{\newblock}{\relax}
\providecommand{\bibinfo}[2]{#2}
\providecommand{\BIBentrySTDinterwordspacing}{\spaceskip=0pt\relax}
\providecommand{\BIBentryALTinterwordstretchfactor}{4}
\providecommand{\BIBentryALTinterwordspacing}{\spaceskip=\fontdimen2\font plus
\BIBentryALTinterwordstretchfactor\fontdimen3\font minus
  \fontdimen4\font\relax}
\providecommand{\BIBforeignlanguage}[2]{{%
\expandafter\ifx\csname l@#1\endcsname\relax
\typeout{** WARNING: IEEEtran.bst: No hyphenation pattern has been}%
\typeout{** loaded for the language `#1'. Using the pattern for}%
\typeout{** the default language instead.}%
\else
\language=\csname l@#1\endcsname
\fi
#2}}
\providecommand{\BIBdecl}{\relax}
\BIBdecl

\bibitem{ref1}
D.~Evans, ``The internet of things: How the next evolution of the internet is
  changing everything,'' 2011.

\bibitem{ref2}
\BIBentryALTinterwordspacing
(2013) World population clock. [Online]. Available:
  \url{http://www.worldometers.info/world-population/}
\BIBentrySTDinterwordspacing

\bibitem{ref3}
M.~Yun and B.~Yuxin, ``Research on the architecture and key technology of
  internet of things (iot) applied on smart grid,'' in \emph{International
  Conference on Advances in Energy Engineering (ICAEE)}, 2010, pp. 69--72.

\bibitem{ref4}
M.~Jung, C.~Reinisch, and W.~Kastner, ``Integrating building automation systems
  and ipv6 in the internet of things,'' in \emph{Sixth International Conference
  on Innovative Mobile and Internet Services in Ubiquitous Computing (IMIS)},
  2012, pp. 683--688.

\bibitem{ref5}
X.~Jia, Q.~Feng, T.~Fan, and Q.~Lei, ``Rfid technology and its applications in
  internet of things (iot),'' in \emph{2nd International Conference on Consumer
  Electronics, Communications and Networks (CECNet)}, 2012, pp. 1282--1285.

\bibitem{ref6}
L.~Coetzee and J.~Eksteen, ``The internet of things - promise for the future?
  an introduction,'' in \emph{IST-Africa Conference Proceedings}, 2011, pp.
  1--9.

\bibitem{ref7}
C.-W. Lu, S.-C. Li, and Q.~Wu, ``Interconnecting zigbee and 6lowpan wireless
  sensor networks for smart grid applications,'' in \emph{Fifth International
  Conference on Sensing Technology (ICST)}, 2011, pp. 267--272.

\bibitem{ref8}
A.~Kouche, ``Towards a wireless sensor network platform for the internet of
  things: Sprouts wsn platform,'' in \emph{IEEE International Conference on
  Communications (ICC)}, 2012, pp. 632--636.

\bibitem{ref9}
R.~Yerra, A.~Bharathi, P.~Rajalakshmi, and U.~Desai, ``Wsn based power
  monitoring in smart grids,'' in \emph{Seventh International Conference on
  Intelligent Sensors, Sensor Networks and Information Processing (ISSNIP)},
  2011, pp. 401--406.

\bibitem{ref10}
L.~Li, H.~Xiaoguang, C.~Ke, and H.~Ketai, ``The applications of wifi-based
  wireless sensor network in internet of things and smart grid,'' in \emph{6th
  IEEE Conference on Industrial Electronics and Applications (ICIEA)}, 2011,
  pp. 789--793.

\bibitem{ref11}
N.~Glombitza, D.~Pfisterer, and S.~Fischer, ``Ltp: An efficient web service
  transport protocol for resource constrained devices,'' in \emph{7th Annual
  IEEE Communications Society Conference on Sensor Mesh and Ad Hoc
  Communications and Networks (SECON)}, 2010, pp. 1--9.

\bibitem{ref12}
L.~Mainetti, L.~Patrono, and A.~Vilei, ``Evolution of wireless sensor networks
  towards the internet of things: A survey,'' in \emph{19th International
  Conference on Software, Telecommunications and Computer Networks (SoftCOM)},
  2011, pp. 1--6.

\bibitem{ref13}
\BIBentryALTinterwordspacing
(2013) Tinyos. [Online]. Available: \url{http://www.tinyos.net}
\BIBentrySTDinterwordspacing

\bibitem{ref14}
\BIBentryALTinterwordspacing
(2013) Memsic professional kit. [Online]. Available:
  \url{http://www.memsic.com/userfiles/files/Datasheets/WSN/6020-0062-06\_A\_WSN\_Professional\_Series.pdf}
\BIBentrySTDinterwordspacing

\bibitem{ref15}
C.~Perkins and E.~Royer, ``Ad-hoc on-demand distance vector routing,'' in
  \emph{Second IEEE Workshop on Mobile Computing Systems and Applications.
  Proceedings. WMCSA '99.}, 1999, pp. 90--100.

\bibitem{ref16}
K.~Mikhaylov and J.~Tervonen, ``Evaluation of power efficiency for digital
  serial interfaces of microcontrollers,'' in \emph{5th International
  Conference on New Technologies, Mobility and Security (NTMS), 2012}, 2012,
  pp. 1--5.

\end{thebibliography}
\end{document}